\documentclass[11pt]{article}
\usepackage{graphicx} 
\usepackage{geometry}
\usepackage{amsmath, amssymb}
\usepackage{hyperref}
\usepackage{setspace}
\usepackage{graphicx} 
\usepackage{subcaption}
\usepackage{lineno}
\usepackage{float}
\usepackage{algorithm}
\usepackage{multirow}
\usepackage{algpseudocode}

\newcommand{\diag}{\mathrm{diag}}

\newcommand{\IR}{\mathbb{R}}

\usepackage{booktabs} 
\usepackage{siunitx} 
\title{\huge Improve Power of Knockoffs with Annotation Information of Covariates}
\author{Xiangyu Zhang\textsuperscript{1},  Lijun Wang\textsuperscript{2}, Changjun Li\textsuperscript{1},Chen Lin\textsuperscript{1}, Hongyu Zhao\textsuperscript{1*}}
\date{}

\begin{document}

\maketitle

\noindent \textbf{1} Department of Biostatistics, School of Public Health, Yale University, New Haven, Connecticut, United States of America

\noindent \textbf{2} Department of Statistics, School of Mathematical Sciences, Zhejiang University, Hangzhou, Zhejiang, China

\bigskip

\noindent *Correspondence: hongyu.zhao@yale.edu

\begin{abstract}
Genome-wide association studies (GWAS) often find association signals between many genetic variants and traits of interest in a genomic region. Functional annotations of these variants provide valuable prior information that helps prioritize biologically relevant variants and enhances the power to detect causal variants. However, due to substantial correlations among these variants, a critical question is how to rigorously control the false discovery rate while effectively leveraging prior knowledge. We introduce annotation-informed knockoffs (AnnoKn), a knockoff-based method that performs annotation-informed variable selection with strict control of the false discovery rate. AnnoKn integrates the knockoff procedure with adaptive Lasso regression to evaluate the importance of multiple covariates while incorporating functional annotation information within a unified Bayesian framework. To facilitate real-world applications where individual-level data are not accessible, we further extend AnnoKn to operate on summary statistics. Through simulations and real-world applications to GTEx and GWAS datasets, we show that AnnoKn achieves superior power in detecting causal genetic variants compared with existing annotation-informed variable selection methods, while maintaining valid control over false discoveries.
\end{abstract}

\section{Introduction}
Genome-wide association studies (GWASs) have achieved remarkable success in detecting genetic variants associated with complex human traits \cite{stranger2011progress}. Other genetic studies, called transcriptome-wide association studies (TWASs) \cite{gamazon2015gene,gusev2016integrative}, have also revealed the molecular regulatory mechanisms underlying human diseases. However, traditional association studies assess only the marginal association between individual covariate–response pairs, ignoring the substantial correlation among genetic factors within a genomic region \cite{visscher2012five,visscher201710}. In high-dimensional genomic analyses, covariates often exhibit complex correlation structures, such as linkage disequilibrium (LD), which complicates the task of distinguishing true causal signals from correlated non-causal covariates \cite{cooper2011needles}. To overcome this challenge and identify genuine causal signals, we may incorporate external information on covariates or explicitly model their underlying dependence structure \cite{cano2020gwas, lu2016genowap, schaid2018genome}.  

With the rapidly increasing volume and diversity of biomedical data, researchers have been gaining more knowledge about the genetic variants that can facilitate genomic analyses and scientific discoveries. An example is the identification of expression quantitative trait loci (eQTLs), which are genetic variants affecting gene expression levels \cite{pickrell2010understanding,gtex2017genetic}. It has been shown that functional annotations—such as regulatory elements, chromatin accessibility, or transcription factor binding sites—can prioritize genetic variants with true regulatory effects, thereby improving the power and accuracy of eQTL detection \cite{watanabe2017functional, wen2015cross, mudappathi2025reg}. Another example is the analysis of GWAS data, where summary statistics ($p$-values or $Z$-scores) from other populations can be leveraged as annotation information to prioritize risk genetic variants before performing association studies in the target population \cite{lu2017post}. Therefore, there is a critical need to develop statistical procedures that leverage functional annotations to improve power while still ensuring rigorous control of the false discovery rate (FDR) within correlated covariates, thereby maintaining the reliability of annotation-informed discoveries. 

Several inference methods have been proposed to identify informative covariates while guaranteeing FDR control. In the classical setting where valid $p$-values are available for all hypotheses, the Benjamini–Hochberg (BH) procedure provides control of FDR when test statistics are independent or satisfy certain forms of positive dependence \cite{benjamini1995controlling, benjamini2001control}. More recently, the adaptive $p$-value thresholding (AdaPT) method has been developed to construct a family of increasingly stringent rejection thresholds using available prior information on covariates \cite{lei2018adapt}. However, standard $p$-value-based FDR control methods require the knowledge of valid $p$-values and specific assumptions about their dependence \cite{benjamini2010discovering}, which result in two issues. First, $p$-values assess only marginal associations, making the resulting discoveries difficult to interpret: many unimportant covariates correlated with true signals can still exhibit strong marginal correlations \cite{brzyski2017controlling}. Second, although various modifications of the BH procedure have been proposed to control the FDR under general dependency structures, these approaches tend to be conservative, often resulting in reduced statistical power \cite{he2024false}. This issue is particularly pronounced in genetics, where complex correlation structures among covariates, such as single nucleotide polymorphisms (SNPs), are prevalent \cite{barrett2005haploview}.

The knockoff framework avoids the direct use of $p$-values and offers a reliable approach to performing controlled variable selection without making assumptions on the conditional distribution of the response \cite{barber2015controlling, candes2018panning}. By modeling the correlation structure of covariates, model-X knockoffs aims to maximize the discovery of conditionally dependent covariates while strictly controlling the FDR at a nominal level. Building on the knockoff framework, the adaptive knockoff filter (AdaKn) has been proposed to incorporate prior annotation information \cite{ren2023knockoffs}. AdaKn first applies the knockoff framework to construct an initial ordering of hypotheses along a data-driven rejection path. It then leverages prior information to refine this ordering by moving covariates with stronger prior support toward the front of the sequence, thereby increasing the likelihood of discovering true signals earlier while maintaining FDR control. However, AdaKn is essentially a two-stage procedure, where the first stage constructs covariate importance measures without effectively incorporating prior information. This may limit the overall statistical power. In addition, AdaKn is designed for settings where both the response and covariates are fully observed, which limits its practical applicability because only summary statistics of marginal associations between each response-covariate pair are available in many real-world settings rather than individual-level data \cite{kontou2024goldmine, zou2022fine}.

We introduce Annotation-informed Knockoffs (AnnoKn), a knockoff-based framework that adaptively leverages functional annotations to enhance the power of knockoffs while maintaining rigorous FDR control. Unlike two-stage approaches, AnnoKn integrates the knockoff procedure and prior information within a unified Bayesian framework. Leveraging prior knowledge of covariates, such as functional annotations, AnnoKn adaptively learns covariate-specific penalty terms to maximize the posterior likelihood. We further extend AnnoKn to operate on summary statistics together with an estimated matrix of pairwise correlations among covariates, without access to individual-level data. Through extensive simulations and real-data analyses, including eQTL mapping and GWASs of Height and Schizophrenia, we demonstrate that AnnoKn achieves higher power than existing methods in detecting causal genetic variants while preserving valid FDR control. In addition, AnnoKn can effectively handle high-dimensional annotations and automatically filter out non-informative annotations.

\section{Methods}

\subsection{Overview of model-X knockoffs and AdaKn}
Let $y$ represent the response, and $\boldsymbol{X}=(X_{1}, ..., X_{p})$ represent $p$ covariates. To infer which covariates are truly associated with the response, model-X knockoffs \cite{candes2018panning} performs conditional tests against the null hypothesis
\[
H_{0}: X_{j} \perp y | X_{-j}, \quad j = 1,\ldots,p,
\]
where $ X_{-j}$ denotes the collection of all covariates except the $j$th one.

To implement model-X knockoffs, one must first construct knockoff copies, denoted by $\tilde{\boldsymbol{X}}=(\tilde{X}_{1}, ..., \tilde{X}_{p})$, which are random variables designed to satisfy the following two conditions:
\begin{itemize}
    \item[] 1. $(\boldsymbol{X}, \tilde{\boldsymbol{X}})_{\text{swap}(S)}\stackrel{\mathrm{d}}{=}(\boldsymbol{X}, \tilde{\boldsymbol{X}}),\  \forall \ \mathrm{subset} \  S \subset \{1,\dots, p\}.$
    \item[] 2. $\tilde{\boldsymbol{X}} \perp y | \boldsymbol{X}.$
\end{itemize}
With knockoff copies, a feature importance statistic $W_j$ is calculated for each covariate. These feature statistics are designed to satisfy two key properties: (1) swapping the $j$th covariate with its knockoff copy has the effect of changing the sign of $W_j$ (the flip sign property); (2) $W_j$ of a non-null covariate tends to take a large positive value. Given a fixed threshold $t>0$, covariates with $W_j \geq t$ are selected as significant. By the flip sign property, the statitics $W_j$ corresponding to null covariates are symmetric around zero, which implies  
\[\#\{j: W_j \leq -t\} \geq \#\{\mathrm{null} \ j: W_j \leq -t\} \stackrel{\mathrm{d}}{=} \#\{\mathrm{null} \ j: W_j \geq t\}.\]
With this inequality, the false discovery proportion (FDP) can be estimated by
\[
\mathrm{FDP}(t) = \frac{\#\{\mathrm{null} \ j: W_j \geq t\}}{\#\{j: W_j \geq t\}} \leq \frac{\#\{j: W_j \leq -t\}}{\#\{j: W_j \geq t\}},
\]
and we can choose the threshold of significant feature statistics $T$ as
\[
T = \min \left\{ t : \frac{1 + \#\{j : W_j \leq -t\}}{\#\{j : W_j \geq t\} \vee 1} \leq q \right\},
\]
where $q$ is the target FDR level. For each covariate $X_{i}$, the $q$-value $q_{i}$ associated with its feature statistic $W_{i}$ is defined as:
\[q_{i} = \min_{t\leq T_{i}} \frac{1 + \#\{j : W_j \leq -t\}}{\#\{j : W_j \geq t\} \vee 1}.\]
The process of selecting covariates while controlling the FDR at a target level $q$ is achieved by identifying covariates for which their corresponding $q$-value, $q_{i}$, satisfies $q_{i} < q$.

The variable selection procedure of model-X knockoffs operates as follows. First, it orders the covariates by sorting the magnitudes of their feature statistics in a non-decreasing manner: 
\[|W_{\pi_1}| \le \dots \le |W_{\pi_k}| \le \dots |W_{\pi_p}|.\]
For each step $k \in\{0, 1,\dots, p\}$, starting from $k = 0$, the method considers selecting all hypotheses $\pi_j$ with $j>k$ and $W_{\pi_j} > 0$, and computes the corresponding estimate of the FDP. The procedure stops at the smallest $k$ for which the estimated FDP falls below the target FDR level.

Given this ordered sequence of hypotheses, AdaKn \cite{ren2023knockoffs} adjusts the ordering by moving covariates with stronger prior support toward the end of the sequence, thereby increasing their likelihood of being selected. At each step $k$, AdaKn uses an adaptive knockoff filter $\phi_{k+1}$ to determine the next hypothesis $\pi_{k+1}$ in the sequence. The FDR control can be demonstrated to hold as long as $\phi_{k+1}$ is fully determined by $\{|W_j|\}_{j \in \{1,\dots,p\}}, \{W_{\pi_j}\}_{j > k}$, and the available prior information. 

AdaKn implements this adaptive knockoff framework using several adaptive filters, including Generalized Linear Model (GLM), Generalized Additive Model (GAM), Random Forest (RF), and Bayesian two-group model (EM) filters, all of which are included in our comparisons.

\subsection{FDR control with knockoffs and adaptive Lasso feature statistics}
Instead of relying on the two-step procedure described above to incorporate prior information, we adopt an adaptive Lasso regression \cite{zou2006adaptive} that computes feature statistics quantifying covariate importance and prioritizes them using annotation information within a unified framework:
\begin{align}
\label{adaptivelasso}
\min_{\boldsymbol{\beta} \in \mathbb{R}^{2p}} l(y,[\boldsymbol{X}, \tilde{\boldsymbol{X}}];\boldsymbol{\beta}) = 
\min_{\boldsymbol{\beta} \in \mathbb{R}^{2p}} \frac{1}{2} \| y - [\boldsymbol{X}, \tilde{\boldsymbol{X}}] \boldsymbol{\beta} \|_2^2 + \lambda \sum_{j = 1}^{p} \phi_{j}(|\beta_{j}|+|\beta_{j+p}|),    
\end{align}
where $\boldsymbol{\beta} = (\beta_{1}, \dots, \beta_{p}, \beta_{p+1}, \dots, \beta_{2p})^{T}$ is the coefficient vector (with the first $p$ components corresponding to the original covariates and the last $p$ to the knockoffs), $\lambda$ is the regularization hyperparameter, and $\phi_{j}$ is the prior importance weight of the $j$-th covariate.

Denoting the solution of the adaptive Lasso problem by $\hat{\boldsymbol{\beta}}$, and defining the coefficient magnitudes as $Z_{j} = |\hat{\beta_{j}}|$ and $\tilde{Z}_{j} = |\hat{\beta}_{j+p}|$, we construct the Lasso Coefficient-Difference (LCD) feature statistic $W_j$ for the $j$-th covariate as:
\[
W_{j}([\boldsymbol{X}, \tilde{\boldsymbol{X}}],y) 
 = Z_{j} - \tilde{Z}_{j} = |\hat{\beta_{j}}|-|\hat{\beta}_{j+p}|.
\]
As noted in the literature \cite{candes2018panning}, valid FDR control is ensured if the feature statistic $W_j$ satisfies the flip sign property. We verify this property for our proposed statistic in Theorem 1.

\noindent
\newline
\textbf{Theorem 1 (Flip-Sign Property of Adaptive Lasso LCD Statistic).} Let $W_{j}$ be the LCD statistic derived from the adaptive Lasso regression defined in (\ref{adaptivelasso}). For any subset of indices $S\subset \{1,...,p\}$, the following flip sign property holds:
\[
w_j \big( [\boldsymbol{X}, \tilde{\boldsymbol{X}}]_{\text{swap}(S)}, y \big) =
\begin{cases}
    w_j \big( [\boldsymbol{X}, \tilde{\boldsymbol{X}}], y \big), & j \notin S, \\
    -w_j \big( [\boldsymbol{X}, \tilde{\boldsymbol{X}}], y \big), & j \in S,
\end{cases}
\]
where $[\boldsymbol{X}, \tilde{\boldsymbol{X}}]_{\text{swap}(S)}$ is obtained from $[\boldsymbol{X}, \tilde{\boldsymbol{X}}]$ by swapping entries $X_{j}$ and $\tilde{X}_{j}$ for every index $j \in S$.

The proof of Theorem 1 can be found in the Supplementary Materials Section S3.1.

\subsection{Annotation-informed knockoffs with individual-level data}
Suppose that we are given $n$ i.i.d. observations $(X_{i1},...,X_{ip},y_{i})$. Let $\mathbf{y} = [y_1,\ldots, y_n]^\top$ be the observed responses for $n$ samples and $\mathbf{X}\in \IR^{n\times p}$ denote the design matrix. Let $\tilde{\mathbf{X}}$ be the constructed knockoff matrix. In this work, we assume that the response vector $\mathbf{y}$, the design matrix $\mathbf{X}$, and the knockoff matrix $\tilde{\mathbf{X}}$ have been standardized to have zero mean and unit variance across all covariates (columns). To define the annotation-informed penalty weights $\phi_j$ for the Adaptive Lasso, we embed the linear regression model within the following Bayesian hierarchical framework:
\begin{align*}
\mathbf{y} = [\mathbf{X}, \tilde{\mathbf{X}}] \boldsymbol{\beta} + \boldsymbol{\epsilon}, \quad \boldsymbol{\epsilon} \sim \mathcal{N}(0, \sigma^2 \mathbf{I}_{n}),
\end{align*}
\begin{align*}
\beta_{j} = \beta_{j+p} \sim Laplace(0, (n\lambda_{0}\phi_{j})^{-1}), \quad j = 1,...,p,
\end{align*}
\begin{align}
\label{annokn}
\phi_{j} = \exp\left(\sum_{l=1}^{L}\lambda_{l}A_{jl}/d\right), \quad \lambda_{l} \sim \mathcal{N}(0, \tau^2),
\end{align}
where $\mathbf{A} = \{A_{jl}\}$ is the standardized $p\times L$ annotation matrix, $\lambda_{0}$ is a pre-selected overall regularization parameter, $\sigma^2$ is the error variance hyperparameter, $d$ controls the scaling or convergence speed of the annotation weights $\lambda_{l}$, and $\tau^2$ is the prior variance hyperparameter for the annotation weights $\lambda_{l}$. 

Denote $\boldsymbol{\lambda} = (\lambda_{1},...,\lambda_{L})$ as the vector of target parameters, $\boldsymbol{\theta} = (\lambda_{0}, \sigma^{2}, d, \tau^{2})$ as the vector of hyperparameters, then the posterior probability can be calculated as:
\begin{align*}
& p(\boldsymbol{\beta}, \boldsymbol{\lambda} |\mathbf{X}, \tilde{\mathbf{X}}, \mathbf{A}, \mathbf{y}, \boldsymbol{\theta}) \\
\propto & p(\mathbf{y}|\mathbf{X}, \tilde{\mathbf{X}},\boldsymbol{\beta},\boldsymbol{\theta}) \times p(\boldsymbol{\beta}|\mathbf{A},\boldsymbol{\lambda},\boldsymbol{\theta}) \times p(\boldsymbol{\lambda}|\boldsymbol{\theta}) \\
= & (2\pi \sigma^{2})^{-n/2} \exp\left(-\frac{1}{2\sigma^2}\|\mathbf{y}-[\mathbf{X}, \tilde{\mathbf{X}}]\boldsymbol{\beta}\|^2\right) \\
& \times \prod_{j=1}^{p} \left(\frac{n\lambda_{0}\phi_{j}}{2}
\right) \exp(-n\lambda_{0}\phi_{j}|\beta_{j}|) \times \prod_{j=1}^{p} \left(\frac{n\lambda_{0}\phi_{j}}{2}\right) \exp(-n\lambda_{0}\phi_{j}|\beta_{j+p}|) \\
& \times (2\pi \tau^{2})^{-L/2} \exp\left(-\frac{1}{2\tau^2}\sum_{l=1}^{L}\lambda_{l}^2\right) \\
\propto & \exp(-\frac{1}{2\sigma^2}\|\mathbf{y}-[\mathbf{X}, \tilde{\mathbf{X}}]\boldsymbol{\beta}\|_2^2) \times \prod_{j=1}^{p} \phi_{j}^{2} \exp(-n\lambda_{0}\phi_{j}(|\beta_{j}| + |\beta_{j+p}|))\times \exp\left(-\frac{1}{2\tau^2}\sum_{l=1}^{L}\lambda_{l}^2\right).
\end{align*}
Given the standardization of $[\mathbf{X}, \tilde{\mathbf{X}}]$ and $\mathbf{y}$ to have unit variance, we fix the error variance hyperparameter at $\sigma^2 = 1$ instead of estimating it. This simplification is justified for two reasons. First, in the context of polygenic traits, the majority of the phenotypic variance is attributable to random error rather than genetic effects (residual variance $\approx 1$) when we restrict analysis to a genomic region. Second, given the structure of the objective function, $\sigma^2$ can effectively be absorbed into the regularization parameters $\lambda_0$ and $\tau^2$. Consequently, fixing $\sigma^2$ does not affect the optimization path or the selection of the target parameters $\boldsymbol{\beta}$ and $\boldsymbol{\lambda}$. The resulting log posterior is therefore:
\begin{align}
\label{loglikelihood}
& \log p(\boldsymbol{\beta}, \boldsymbol{\lambda} |\mathbf{X}, \tilde{\mathbf{X}}, \mathbf{A}, \mathbf{y}, \boldsymbol{\theta}) \nonumber\\
= & -\frac{1}{2}\|\mathbf{y}-[\mathbf{X}, \tilde{\mathbf{X}}]\boldsymbol{\beta}\|_2^2 + 2 \sum_{j=1}^{p} \log \phi_{j} -n\lambda_{0} \sum_{j=1}^{p} \phi_{j}(|\beta_{j}| + |\beta_{j+p}|) -\frac{1}{2\tau^2}\sum_{l=1}^{L}\lambda_{l}^2 + \text{const}.
\end{align}
Note that when $\boldsymbol{\lambda}$ is fixed, maximizing the log posterior in (\ref{loglikelihood}) with respect to $\boldsymbol{\beta}$ is equivalent to solving the adaptive Lasso model in (\ref{adaptivelasso}). Consequently, we propose an alternating maximization strategy to optimize the target log posterior with respect to $\boldsymbol{\beta}$ and $\boldsymbol{\lambda}$, given a fixed $\lambda_0$:

\begin{itemize}
    \item Step 1: We initialize the prior weights $\phi_{j} = 1$ for $j = 1,\dots,p$, and set the initial $\boldsymbol{\beta}$ as the Ordinary Least Squares (OLS) estimates.
    \item Step 2: Given current estimates of $\phi_{j}$, we update $\hat{\boldsymbol{\beta}}$ by solving the adaptive Lasso problem:  \[\hat{\boldsymbol{\beta}} = \arg\min_{\boldsymbol{\beta}} \left\{ \frac{1}{2n}\|\mathbf{y}-[\mathbf{X}, \tilde{\mathbf{X}}]\boldsymbol{\beta}\|_2^2 + \lambda_{0} \sum_{j = 1}^{p} \phi_{j}(|\beta_{j}|+|\beta_{j+p}|) \right\}.\]
    \item Step 3: Given current estimates of $\boldsymbol{\beta}$, and leveraging the fact that $\mathbf{A}$ is standardized such that $\sum_{j=1}^{p} A_{jl} = 0$ for $l \in \{1,\dots,L\}$, we update $\hat{\boldsymbol{\lambda}}$ by maximizing the following reduced expression with respect to $\boldsymbol{\lambda}$:
    \begin{align}
    \label{maxlambda}
    & 2 \sum_{j=1}^{p} \log \phi_{j} -n\lambda_{0} \sum_{j=1}^{p} \phi_{j}(|\beta_{j}| + |\beta_{j+p}|) -\frac{1}{2\tau^2}\sum_{l=1}^{L}\lambda_{l}^2 \nonumber \\
    = & 2 \sum_{j=1}^{p} \sum_{l=1}^{L}\lambda_{l}A_{jl}/d -n\lambda_{0} \sum_{j=1}^{p} \exp(\sum_{l=1}^{L}\lambda_{l}A_{jl}/d)(|\beta_{j}| + |\beta_{j+p}|) -\frac{1}{2\tau^2}\sum_{l=1}^{L}\lambda_{l}^2 \nonumber \\
    = & -n\lambda_{0} \sum_{j=1}^{p} \exp(\sum_{l=1}^{L}\lambda_{l}A_{jl}/d)(|\beta_{j}| + |\beta_{j+p}|) -\frac{1}{2\tau^2}\sum_{l=1}^{L}\lambda_{l}^2,
    \end{align}
    and update $\phi_{j}$ based on updated estimates of $\boldsymbol{\lambda}$.

    \item Step 4 (Iteration): Iterate between Step 2 and Step 3 until convergence is achieved, and output the final estimates $\hat{\boldsymbol{\beta}}$ and $\hat{\boldsymbol{\lambda}}$.

\end{itemize}

Building upon this strategy, we propose Algorithm \ref{knockoff_anno}, which tunes the hyperparameter $\lambda_0$ among a candidate set $\mathbb{A}$ and selects the best-performing $\lambda_0$ to minimize the cross-validation error.

\begin{algorithm}[ht]
\caption{Main algorithm of annotation-informed knockoffs (AnnoKn)}
\label{knockoff_anno}
\begin{algorithmic}[1]
\Require response vector $\mathbf{y}$, covariate matrix $\mathbf{X}$, knockoff copy $\tilde{\mathbf{X}}$, annotation matrix $\mathbf{A}$, hyperparameters $d, \tau^{2}$
\State \textbf{Initialize:} $\lambda^{f}_{0}$ = 0, $c^{f} = \infty$, $\phi^{f}_{j} = 1$ for $j = 1, \dots, p$
\For{$\lambda_0$ in $\mathbb{A}$}
\State \textbf{Initialize:} set $\phi_j = 1$ for $j = 1, \dots, p$, and $\boldsymbol{\beta}$ as the OLS estimates
\Repeat
    \State Given current $\phi_j$, estimate $\boldsymbol{\beta}$ by solving:
    \[
    \min_{\boldsymbol{\beta}} \left\{
    \frac{1}{2n} \| \mathbf{y} - [\mathbf{X}, \tilde{\mathbf{X}}] \boldsymbol{\beta} \|_2^2 +
    \lambda_0 \sum_{j=1}^p \phi_j (|\beta_j| + |\beta_{j+p}|)
    \right\}
    \]
    \State Given current $\boldsymbol{\beta}$, update $\boldsymbol{\lambda}$ by maximizing:
    \[
    -n\lambda_{0} \sum_{j=1}^{p} \exp\left(\sum_{l=1}^{L}\lambda_{l}A_{jl}/d\right)(|\beta_{j}| + |\beta_{j+p}|) -\frac{1}{2\tau^2}\sum_{l=1}^{L}\lambda_{l}^2
    \]
    \State Update $\phi_{j} = \exp\left(\sum_{l=1}^{L}\lambda_{l}A_{jl}/d\right)$ for $j = 1, \dots, p$
\Until{convergence of $\boldsymbol{\beta}$ and $\boldsymbol{\lambda}$}
\State \textbf{Cross-validation:} calculate cross-validation error $c$ with current $\lambda_0$ and $\phi_{j}$
\If{$c < c^{f}$}
  \State Update $\lambda^{f}_{0} \leftarrow \lambda_{0}$, $c^{f} \leftarrow c$, and $\phi^{f}_{j} \leftarrow \phi_{j}$ for $j = 1, \dots, p$
\EndIf
\EndFor
\State \textbf{Final fit:} apply adaptive Lasso with $\{\phi^{f}_{j}\}_{j=1}^p$ and $\lambda^{f}_0$ to estimate $\boldsymbol{\beta}$
\State \textbf{Output:} final estimates of $\boldsymbol{\beta}$, $\{\phi^{f}_{j}\}_{j=1}^p$ and $\lambda^{f}_0$
\end{algorithmic}
\end{algorithm}

Although Algorithm \ref{knockoff_anno} achieves good performance, it can be computationally intensive due to its dependence on the size of the candidate set $\mathbb{A}$ for $\lambda_0$. We therefore propose Algorithm \ref{knockoff_simple}, which has been demonstrated to have similar performance to Algorithm \ref{knockoff_anno} while offering substantially higher computational efficiency.

\begin{algorithm}[ht]
\caption{Simplified version of annotation-informed knockoffs (AnnoKn-lite)}
\label{knockoff_simple}
\begin{algorithmic}[1]
\Require response vector $\mathbf{y}$, covariate matrix $\mathbf{X}$, knockoff copy $\tilde{\mathbf{X}}$, annotation matrix $\mathbf{A}$, hyperparameters $d, \tau^{2}$
\State \textbf{Initialize:} set $\phi_j = 1$ for $j = 1, \dots, p$, and select the optimal $\lambda_0$ using cross-validation on the standard Lasso problem
\Repeat
    \State Given current $\phi_j$, estimate $\boldsymbol{\beta}$ by solving:
    \[
    \min_{\boldsymbol{\beta}} \left\{
    \frac{1}{2n} \| \mathbf{y} - [\mathbf{X}, \tilde{\mathbf{X}}] \boldsymbol{\beta} \|_2^2 +
    \lambda_0 \sum_{j=1}^p \phi_j (|\beta_j| + |\beta_{j+p}|)
    \right\}
    \]
    \State Given current $\boldsymbol{\beta}$, update $\boldsymbol{\lambda}$ by maximizing:
    \[
     -n\lambda_{0} \sum_{j=1}^{p} \exp\left(\sum_{l=1}^{L}\lambda_{l}A_{jl}/d\right)(|\beta_{j}| + |\beta_{j+p}|) -\frac{1}{2\tau^2}\sum_{l=1}^{L}\lambda_{l}^2
    \]
    \State Update $\phi_{j} = \exp\left(\sum_{l=1}^{L}\lambda_{l}A_{jl}/d\right)$ for $j = 1, \dots, p$
\Until{convergence of $\boldsymbol{\beta}$ and $\boldsymbol{\lambda}$}
\State \textbf{Re-tune:} with final estimates of $\{\phi_j\}_{j=1}^p$, re-tune $\lambda_0$ with cross-validation         
\State \textbf{Final fit:} apply adaptive Lasso with $\{\phi_j\}_{j=1}^p$ and $\lambda_0$ to estimate $\boldsymbol{\beta}$
\State \textbf{Output:} final estimates of $\boldsymbol{\beta}$, $\{\phi_j\}_{j=1}^p$ and $\lambda_0$
\end{algorithmic}
\end{algorithm}

\subsection{Annotation-informed knockoffs with summary statistics}
Consider a general case where we relate the response vector $\mathbf{y}$ with the design matrix $\mathbf{X} =  (\mathbf{x}_{1}, \dots, \mathbf{x}_{p})$ using the multiple linear regression model:
$$\mathbf{y}_{n\times 1} = \mathbf{X}_{n\times p}\boldsymbol{b}_{p\times 1} + \boldsymbol{\epsilon}_{n\times 1}, \quad \boldsymbol{\epsilon} \sim \mathcal{N}(0, \sigma^2I_{n}),$$
where both $\mathbf{y}$ and $\mathbf{X}$ have been standardized to have zero mean and unit variance. The vector $\boldsymbol{b}$ denotes the true effect sizes of the $p$ covariates. 

In practice, accessing individual-level data, which comprises the phenotype vector $\mathbf{y}$ and the genotype matrix $\mathbf{X}$, may be challenging for most GWASs due to privacy constraints. Instead, these studies usually provide summary statistics, which capture the marginal association evidence for each covariate $j$.

The summary statistics include the estimated marginal effect size ($\hat{b}_{j}$) and the corresponding standard error ($se(\hat{b}_{j})$), often derived from a simple linear regression model run for each covariate separately:
$$\mathbf{y} = \mathbf{x}_{j}{b}_{j} + \boldsymbol{\epsilon}_{j}, \quad \boldsymbol{\epsilon}_{j} \sim \mathcal{N}(0, \sigma_{j}^2).$$
The Knockoff framework has been extended to operate entirely on summary statistics. Specifically, the GhostKnockoff method \cite{he2022ghostknockoff, chen2024controlled} performs controlled variable selection by leveraging only the marginal $Z$-scores and an estimated correlation matrix among covariates, that can be obtained from an external data source, such as a reference panel from the 1000 Genomes Project (1KG).

Denote $\hat{\boldsymbol{b}} = (\hat{b}_{1},\dots,\hat{b}_{p})^\top$ as the vector of effect sizes estimated from the marginal linear regressions. Under the assumption that each genetic variant has only a small correlation with the phenotype $\mathbf{y}$, it was shown in \cite{zhu2017bayesian} that
\begin{equation}
\label{likelihood1}
\hat{\boldsymbol{b}}|\boldsymbol{b}, \boldsymbol{S}, \boldsymbol{\Sigma} \sim \mathcal{N}\left(\boldsymbol{S}\boldsymbol{\Sigma}\boldsymbol{S}^{-1}\boldsymbol{b}, \boldsymbol{S}\boldsymbol{\Sigma}\boldsymbol{S}\right),
\end{equation}
where $\boldsymbol{S} = \diag(se(\hat{b}_{1}),\dots, se(\hat{b}_{p}))$, $\boldsymbol{\Sigma}$ is the LD matrix representing the pairwise correlations among genotypes of $p$ SNPs. 

Denote $z_{j} = \hat{b}_{j}/se(\hat{b}_{j})$ as the marginal $Z$-score for SNP $j$, $\boldsymbol{z} = (z_{1},\dots,z_{p})^{\top}$ is the vector of marginal $Z$-scores. The conditional distribution of $\boldsymbol{z} = \boldsymbol{S}^{-1}\hat{\boldsymbol{b}}$ is
\begin{equation}
\label{likelihood2}
\boldsymbol{z}|\boldsymbol{b}, \boldsymbol{S}, \boldsymbol{\Sigma} \sim \mathcal{N}\left(\boldsymbol{\Sigma}\boldsymbol{S}^{-1}\boldsymbol{b}, \boldsymbol{\Sigma}\right).
\end{equation}
For polygenic traits, the phenotypic variance explained by any individual SNP is typically negligible, allowing us to assume $\sigma^2_{j} \approx 1$. Moreover, since each genotype vector $\mathbf{x}_{j}$ has been standardized, we have $\mathbf{x}^{\top}_{j}\mathbf{x}_{j} = n$. Therefore, the standard error of $\hat{b}_{j}$ can be calculated as
$$se(\hat{b}_{j}) = \sqrt{\sigma_{j}^2/\mathbf{x}^{\top}_{j}\mathbf{x}_{j}} \approx \sqrt{1/n}.$$
With this approximation, the conditional distributions in (\ref{likelihood1} and \ref{likelihood2}) simplify to
$$\hat{\boldsymbol{b}}|\boldsymbol{b}, \boldsymbol{\Sigma} \sim \mathcal{N}\left(\boldsymbol{\Sigma}\boldsymbol{b}, \boldsymbol{\Sigma}/n\right),$$ 
$$\boldsymbol{z}|\boldsymbol{b}, \boldsymbol{\Sigma} \sim \mathcal{N}\left(\sqrt{n}\boldsymbol{\Sigma}\boldsymbol{b}, \boldsymbol{\Sigma}\right).$$
Given this conditional distribution of $\boldsymbol{z}$, we can extend AnnoKn to operate using only summary statistics together with an estimated LD matrix $\boldsymbol{\Sigma}$, which can be obtained from an external reference panel. We adopt a multiple-knockoff framework where the z-scores of $M$ knockoff copies, $(\tilde{\boldsymbol{z}}_{1},\dots, \tilde{\boldsymbol{z}}_{M})$, are generated simultaneously. Following \cite{gimenez2019improving}, the joint correlation structure between the original covariates and the $M$ knockoff copies takes the form:
\[\boldsymbol{\Sigma}_{M} = 
\underbrace{
    \begin{pmatrix}
        \boldsymbol{\Sigma} & \boldsymbol{\Sigma} - \mathbf{D} & \dots & \boldsymbol{\Sigma} - \mathbf{D} \\
        \boldsymbol{\Sigma} - \mathbf{D} & \boldsymbol{\Sigma} & \dots & \boldsymbol{\Sigma} - \mathbf{D} \\
        \vdots & \vdots & \ddots & \vdots \\
        \boldsymbol{\Sigma} - \mathbf{D} & \boldsymbol{\Sigma} - \mathbf{D} & \dots & \boldsymbol{\Sigma}
    \end{pmatrix},
}_{M+1 \text{ blocks}}\]
where $\mathbf{D} = \diag(s_1,\dots,s_p)$ is a diagonal matrix given by solving the following convex optimization problem:
\[\text{minimize} \quad \sum_j |1 - s_j|, \quad \text{subject to} \quad 
\begin{cases} 
    \frac{M+1}{M}\boldsymbol{\Sigma} - \mathbf{D} \succeq 0, \\
    s_j \ge 0, 1 \le j \le p.
\end{cases}\]
Denote $\boldsymbol{Z}_{M} = (\boldsymbol{z}^\top, \tilde{\boldsymbol{z}}_1^\top, \dots, \tilde{\boldsymbol{z}}_M^\top)^\top$ as an $(M+1)p$ vector containing the combined marginal $Z$-scores for the original covariates and $M$ knockoff copies, then the conditional distribution of $\boldsymbol{Z}_{M}$ is:
\[\boldsymbol{Z}_{M}|\boldsymbol{\beta}, n, \boldsymbol{\Sigma}_{M} \sim \mathcal{N}\left(\sqrt{n}\boldsymbol{\Sigma}_{M}\boldsymbol{\beta}, \boldsymbol{\Sigma}_{M}\right),\]
where $\boldsymbol{\beta} = (\beta_{1},\dots,\beta_{(M+1)p})^\top$ is the vector of true coefficients for the original covariates and $M$ knockoff copies.

Based on this conditional distribution, we perform annotation-informed GhostKnockoff (AnnoGK) using summary statistics and an estimated LD matrix within the following Bayesian framework. Similar to AnnoKn, we also place Laplace priors for the coefficients:
\begin{align*}
\boldsymbol{Z}_{M}|\boldsymbol{\beta}, n, \boldsymbol{\Sigma}_{M} \sim \mathcal{N}\left(\sqrt{n} \boldsymbol{\Sigma}_{M}\boldsymbol{\beta}, \boldsymbol{\Sigma}_{M}\right),
\end{align*}
\begin{align*}
\beta_{j+k\times p} \sim Laplace(0, (n\lambda_{0}\phi_{j})^{-1}), \quad k = 0,\dots,M, \quad j = 1,\dots,p,
\end{align*}
\begin{align*}
\phi_{j} = \exp\left(\sum_{l=1}^{L}\lambda_{l}A_{jl}/d\right), \quad \lambda_{l} \sim \mathcal{N}(0, \tau^2),
\end{align*}
where $n$ is the sample size. We can then calculate the posterior distribution of target parameters $\boldsymbol{\beta}$ and $\boldsymbol{\lambda}$ as
\begin{align*}
& p(\boldsymbol{\beta}, \boldsymbol{\lambda} |\boldsymbol{Z}_{M}, \boldsymbol{\Sigma}_{M}, n, \mathbf{A}, \boldsymbol{\theta}) \\
\propto & p(\boldsymbol{Z}_{M}|\boldsymbol{\beta},\boldsymbol{\Sigma}_{M}, n,\boldsymbol{\theta}) \times p(\boldsymbol{\beta}|\mathbf{A},\boldsymbol{\lambda},\boldsymbol{\theta}) \times p(\boldsymbol{\lambda}|\boldsymbol{\theta}) \\
\propto & \exp\left(-\frac{1}{2}(\boldsymbol{Z}_{M} - \sqrt{n}\boldsymbol{\Sigma}_{M}\boldsymbol{\beta})^\top \boldsymbol{\Sigma}_{M}^{-1}(\boldsymbol{Z}_{M} - \sqrt{n}\boldsymbol{\Sigma}_{M}\boldsymbol{\beta})\right) \\
& \times \prod_{j=1}^{p} \prod_{k=0}^{M}\left(\frac{n\lambda_{0}\phi_{j}}{2}\right) \exp(-n\lambda_{0}\phi_{j}|\beta_{j+k\times p}|)  \times (2\pi \tau^{2})^{-L/2} \exp\left(-\frac{1}{2\tau^2}\sum_{l=1}^{L}\lambda_{l}^2\right).
\end{align*}
The log-likelihood is 
\begin{align*}
& \log p(\boldsymbol{\beta}, \boldsymbol{\lambda} |\boldsymbol{Z}_{M}, \boldsymbol{\Sigma}_{M}, n, \mathbf{A}, \boldsymbol{\theta}) \\
= & -\frac{1}{2}(\boldsymbol{Z}_{M} - \sqrt{n}\boldsymbol{\Sigma}_{M}\boldsymbol{\beta})^\top \boldsymbol{\Sigma}_{M}^{-1}(\boldsymbol{Z}_{M} - \sqrt{n}\boldsymbol{\Sigma}_{M}\boldsymbol{\beta}) + (M+1)\sum_{j = 1}^{p}\log \phi_{j}  \\
& -n \lambda_0 \sum_{j = 1}^{p}\phi_j\sum_{k = 0}^{M}|\beta_{j+k\times p}| -\frac{1}{2\tau^2}\sum_{l=1}^{L}\lambda_{l}^2 + \text{const} \\
= & -\frac{n}{2}\boldsymbol{\beta}^\top \boldsymbol{\Sigma}_{M} \boldsymbol{\beta} + \sqrt{n}\boldsymbol{\beta}^\top  \boldsymbol{Z}_{M} + (M+1)\sum_{j = 1}^{p}\log \phi_{j} -n \lambda_0 \sum_{j = 1}^{p}\phi_j\sum_{k = 0}^{M}|\beta_{j+k\times p}| - \frac{1}{2\tau^2}\sum_{l=1}^{L}\lambda_{l}^2 + \text{const}.
\end{align*}
Given a fixed hyperparameter $\lambda_0$, we iteratively update
$\boldsymbol{\beta}$ and $\boldsymbol{\lambda}$ to maximize this log-posterior:
\begin{itemize}
    \item Given $\boldsymbol{\lambda}$, update $\boldsymbol{\beta}$ by minimizing the loss function:
    \[
    \frac{1}{2}\boldsymbol{\beta}^\top \boldsymbol{\Sigma}_{M} \boldsymbol{\beta} -\frac{1}{\sqrt{n}} \boldsymbol{\beta}^\top \boldsymbol{Z}_{M} + \lambda_0 \sum_{j = 1}^{p}\phi_j\sum_{k = 0}^{M}|\beta_{j+k\times p}|
    \]
    using the BASIL framework \cite{qian2020fast}.
    \item Given $\boldsymbol{\beta}$, update $\boldsymbol{\lambda}$ by maximizing the objective function:
    \begin{align}
    & (M+1)\sum_{j = 1}^{p}\phi_{j} -n \lambda_0 \sum_{j = 1}^{p}\phi_j(\sum_{k = 0}^{M}|\beta_{j+k\times p}|) - \frac{1}{2\tau^2}\sum_{l=1}^{L}\lambda_{l}^2  \nonumber \\
    = & (M+1) \sum_{j=1}^{p} \sum_{l=1}^{L}\lambda_{l}A_{jl}/d -n\lambda_{0} \sum_{j=1}^{p} \exp(\sum_{l=1}^{L}\lambda_{l}A_{jl}/d)(\sum_{k = 0}^{M}|\beta_{j+k\times p}|) -\frac{1}{2\tau^2}\sum_{l=1}^{L}\lambda_{l}^2 \nonumber \\
    = & -n\lambda_{0} \sum_{j=1}^{p} \exp(\sum_{l=1}^{L}\lambda_{l}A_{jl}/d)(\sum_{k = 0}^{M}|\beta_{j+k\times p}|) -\frac{1}{2\tau^2}\sum_{l=1}^{L}\lambda_{l}^2.
    \end{align}
\end{itemize}

We summarize the algorithm of AnnoGK in Supplementary Materials Section 3.4. To select the optimal value $\lambda_0$, we adopt the pseudo–summary statistics approach proposed in \cite{zhang2021improved} and referenced by GhostKnockoff in \cite{chen2024controlled}. The core idea is to construct training and validation summary statistics using designated training and validation sample sizes $n_{t}$ and $n_{v}$, respectively. For a candidate set $\mathbb{A}$ of possible $\lambda_0$ values, we identify the best-performing value by maximizing an approximate estimate of the correlation between the predicted Lasso coefficients and their true counterparts based on the pseudo-validation summary statistics. In this study, we employ a five-fold cross-validation scheme with $n_t = 0.8n$ and $n_v = 0.2n$. We also fix $M = 1$ throughout the manuscript to focus on validating the fundamental usage of AnnoGK.

\section{Simulation studies for individual-level data}

In this section, we present simulation studies using individual-level data $\mathbf{X}$ and $\mathbf{y}$ to illustrate that AnnoKn improves the power of variable selection by effectively leveraging annotation information while maintaining valid FDR control. Following the simulation setting of AdaKn, we consider two settings: one-dimensional and two-dimensional annotations.

\subsection{One-dimensional annotation information} 
We first replicated the primary simulation scenario from the AdaKn study (details provided in Supplementary Materials Section S4.1). To mimic real genotype data, the covariates were drawn from a hidden Markov model (HMM) \cite{sesia2019gene}. The simulated dataset consisted of $n = 1{,}000$ samples and $p = 900$ covariates. We randomly selected 150 causal covariates from the first 300, such that the probability of selection decreases as the covariate index increases. Each causal covariate was assigned a coefficient $\beta_{j} = \pm 3.5/\sqrt{n}$, while all non-causal covariates had coefficients set to 0. The annotation was defined as the covariate index: $A_j = j$ for $j=1,\dots,p$.

\begin{figure}[htbp]
  \centering
  \includegraphics[width=1\textwidth]{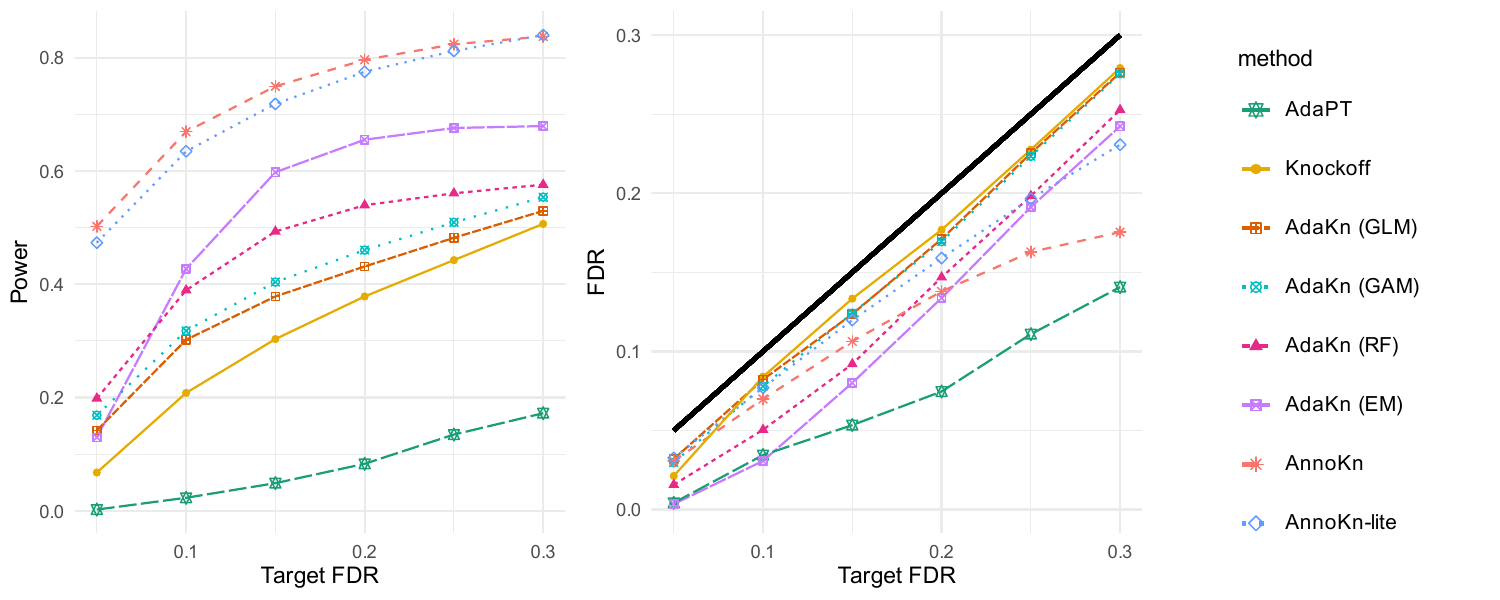}
  \caption{{\bf Comparison of Knockoffs, AnnoKn, AdaPT, and AdaKn with individual-level data and one-dimensional continuous annotation.}  The target $q$-value ranges from 0.05 to 0.3, with the black solid line indicating the target $q$-value level.}
  \label{fig:sim1-1dim}
\end{figure}

We repeated this simulation 100 times and compared AnnoKn with AdaPT, Knockoffs, and AdaKn with multiple types of knockoff filters, including GLM, GAM, random forest (RF), and two-group model (EM). As shown in Figure \ref{fig:sim1-1dim}, AnnoKn and AnnoKn-lite consistently outperformed all AdaKn methods, AdaPT, and original knockoff procedure (which ignored annotation information), achieving substantially higher power while maintaining FDR below the target level. Notably, AnnoKn-lite, despite its simplified structure, attained power slightly lower but nearly identical to AnnoKn.

We also conducted the same simulation after converting the continuous index into a binary annotation. To be more specific, we set $A_j = 0$ for $j\in \{1,\dots,300\}$ and $A_j = 1$ otherwise (Supplemental Figure S1). Although lower than the power achieved using continuous annotation (unsurprising as binary annotation information contains less information), AnnoKn and AnnoKn-lite still outperformed competing methods. AdaPT was not able to incorporate binary annotation, and AdaKn (RF) performed poorly under strict FDR control.

\subsection{Two-dimensional annotation information} 

Following AdaKn's setting, we simulated the dataset with $n = 1,000$ samples and $p = 1,600$ covariates. The $n\times p$ design matrix $\mathbf{X}$ was generated with i.i.d Gaussian entries. Conditional on $\mathbf{X}$, the response vector $\mathbf{y}$ was generated using a logistic model (see Supplementary Materials Section S4.2 for further details). The coefficient vector $\boldsymbol{\beta}$ was structured spatially on a $40\times 40$ grid. To comprehensively evaluate the performance of AnnoKn with other methods, we considered four region types within which the coefficients could be causal (Supplemental Figure S2). Within each causal region, the probability of each coefficient $\beta_{j}$ being nonzero was determined by its corresponding coordinates $\bigl(r(j),\,s(j)\bigr)$ on the plane, and nonzero coefficients were set to $25/\sqrt{n}$, while their signs were determined by i.i.d. coin flips. 

\begin{figure}[htbp]
  \centering
  \includegraphics[width=1\textwidth]{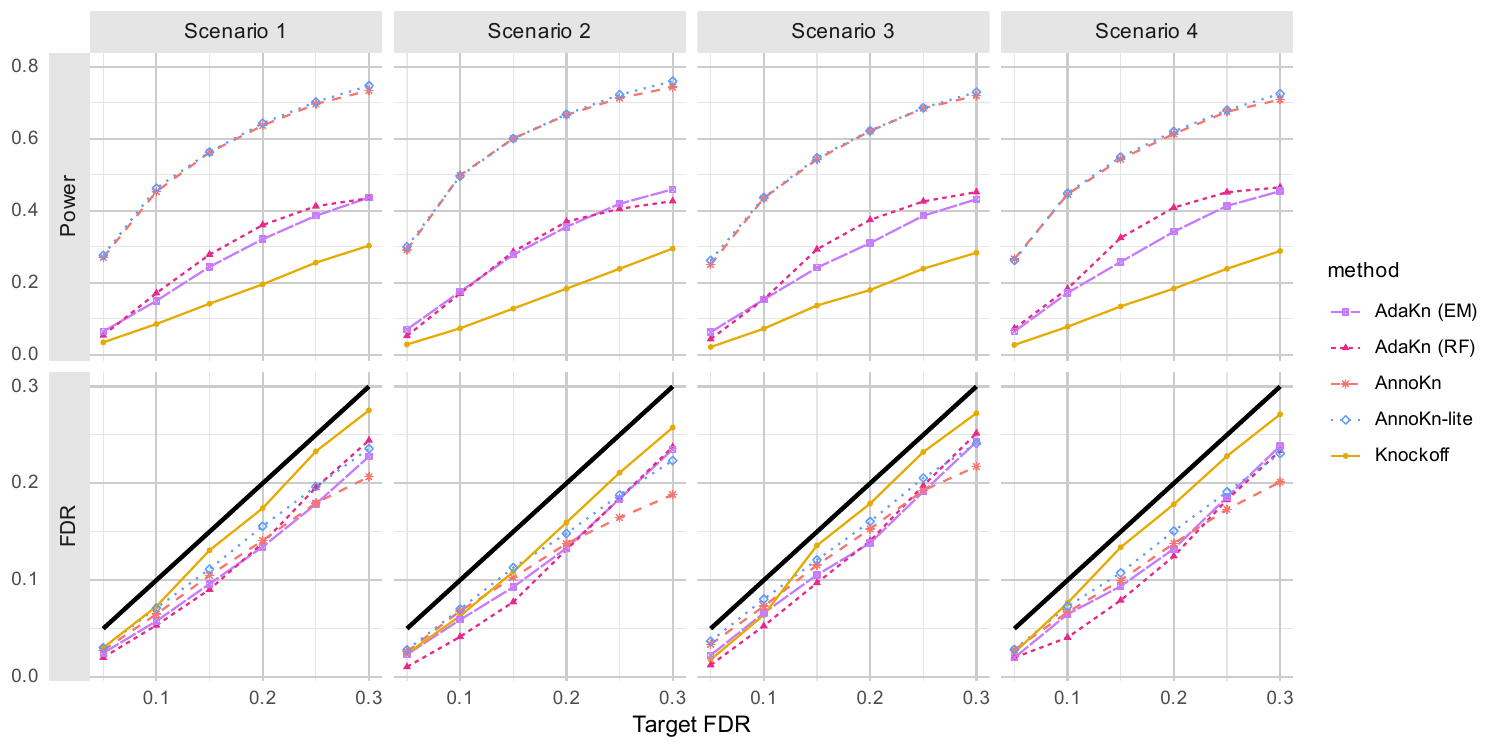}
  \caption{{\bf Comparison of Knockoffs, AnnoKn, and AdaKn with individual-level data and two-dimensional continuous annotations.}  The target $q$-value ranges from 0.05 to 0.3, with the black solid line indicating the theoretical $q$-value level.}
  \label{fig:sim2-2dim}
\end{figure}

We first compared AnnoKn and AnnoKn-lite with AdaKn and Knockoffs using the coordinates $\bigl(r(j),\,s(j)\bigr)$ for each coefficient as its two-dimensional continuous annotations. As shown in Figure \ref{fig:sim2-2dim}, all methods consistently controlled FDR. However, AnnoKn and AnnoKn-lite achieved substantially higher power than Knockoffs and AdaKn variants employing EM and RF filters. While AnnoKn attained similar power to AnnoKn-lite, it yielded lower FDR.

We next evaluated performance given binary annotations. Specifically, we set $r(j) = 0$ for all covariates whose $r$-coordinates did not exceed the maximum $r$-coordinate of the causal region, and $r(j) = 1$ otherwise. The $s(j)$ coordinates were binarized analogously. As shown in Supplemental Figure S3, both AnnoKn and AnnoKn-lite continued to control the FDR and improve statistical power. 

To assess robustness of AnnoKn to non-informative annotation, we extended the one-dimensional annotation setting by adding four additional non-informative annotations created by randomly permuting $\{1,\dots,p\}$. As shown in Supplemental Figures S4 and S6, and discussed in Supplementary Materials Section S5.1, AnnoKn and AnnoKn-lite displayed strong robustness to the inclusion of noise compared with AdaKn. In particular, AnnoKn automatically reduced the influence of non-informative annotations by shrinking their corresponding parameters ($\lambda_l$) towards zero (Supplemental Figures S5 and S7).

\section{Simulation studies based on summary statistics}

To illustrate that integrating annotation information into GhostKnockoff via AnnoGK improves model performance, we conducted numerical simulations under settings similar to those used by GhostKnockoff \cite{chen2024controlled}. We considered three sample size settings: $(n,p) \in \{(5,000, 300), \\ (5,000, 600), (5,000, 1,000)\}$. We generated $\mathbf{x}_{i} \overset{\text{iid}}{\sim} \mathcal{N}(0, \Sigma_{\rho})$ for $i \in \{1,2,...,n\}$, where $[\Sigma_\rho]_{s,t} = \rho^{|s - t|}$ for $1\leq s, t \leq p$. We fixed $\rho = 0.5$ to introduce a moderate correlation among covariates. 

The responses were generated as $y_{i} =\boldsymbol{\beta}^{\top}\mathbf{x}_{i} + \epsilon_{i}$, where $\epsilon_{i} \overset{\text{iid}}{\sim} \mathcal{N}(0, \sigma^{2})$. We created a sparse coefficient vector $\boldsymbol{\beta}$ by selecting 30 coordinates among the first 60 to be nonzero such that the probability of the $j$-th covariate being causal was proportional to $j^{-2}$. The signs of signals with nonzero coefficients were assigned to be either positive or negative with equal probability, with the constant magnitude for each signal and the parameter $\sigma^{2}$ selected to control the variation explained by genetic factors $h^2 := \mathrm{Var}({\boldsymbol{\beta}^{\top}\mathbf{x}})/\mathrm{Var}(\mathbf{y}) \in \{0.05, 0.1, 0.2\}$.

\begin{figure}[htbp]
  \centering
  \includegraphics[width=1\textwidth]{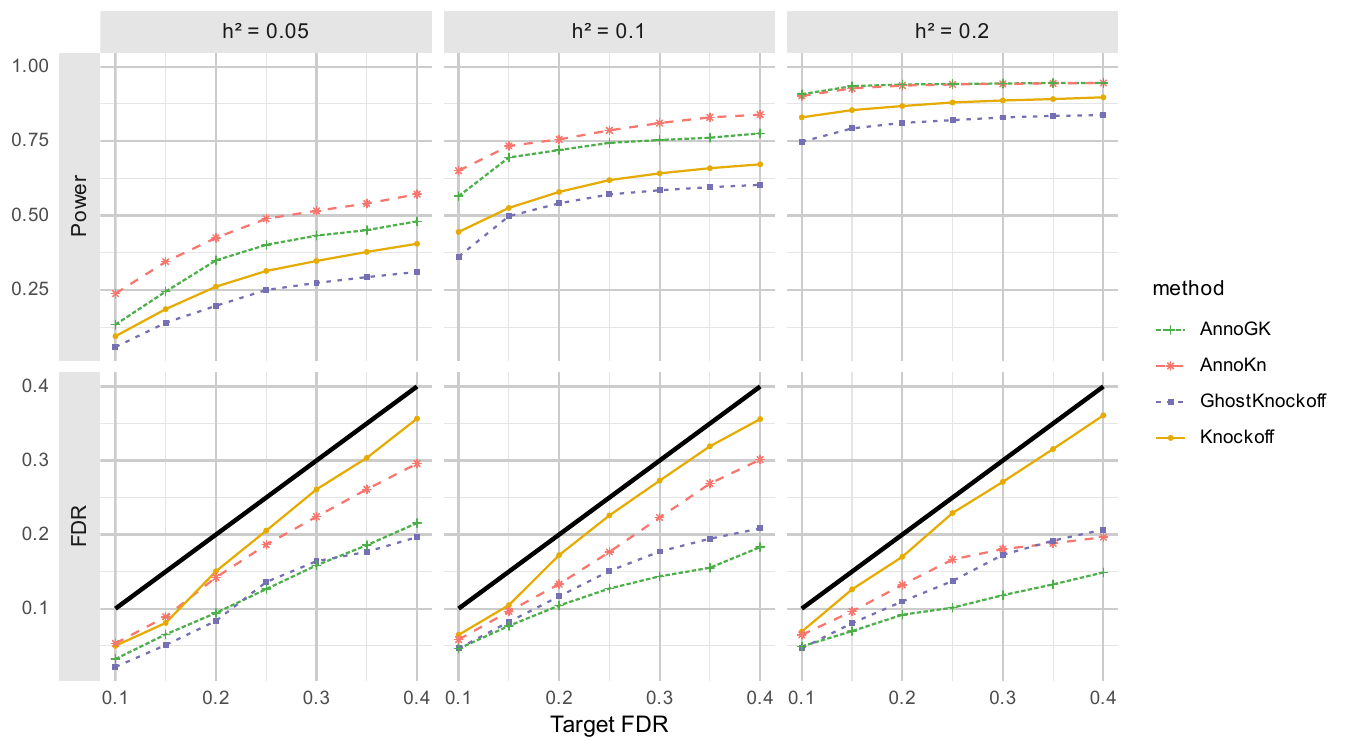}
  \caption{{\bf Comparison of Knockoffs, AnnoKn, GhostKnockoff, and AnnoGK with simulated AR(1) design matrix.}  We vary the phenotypic variance explained by covariates ($h^2$) across $\{0.05, 0.1, 0.2\}$. The $q$-value threshold ranges from 0.1 to 0.4, with the black solid line indicating the theoretical $q$-value level. The number of covariates was $p = 1,000$. We used the index of each covariate as the continuous annotation information.}
  \label{AnnoGK_simu}
\end{figure}

We first considered using the index of each covariate as continuous annotation information. For each setting, we considered four methods: Knockoffs (using the LCD statistic), AnnoKn, GhostKnockoff (using the LCD statistic, with tuning parameter $\lambda_0$ chosen by the pseudo-summary statistics approach), and AnnoGK. Figure \ref{AnnoGK_simu} presents the power and FDR of each method versus the target FDR levels when $p = 1,000$. Across all heritability values, all methods appropriately controlled FDR. The annotation-informed methods (AnnoGK and AnnoKn) consistently achieved higher power and lower realized FDR compared to their respective counterparts, GhostKnockoff and Knockoffs. Summary-statistic-based methods tended to be more conservative with stricter FDR control, while the power of AnnoGK was very close to that of AnnoKn, particularly under stronger signals ($h^2 = 0.2$). This pattern also held for $p \in \{300, 600\}$, as shown in Supplemental Figures S9 and S10, which exhibited nearly identical power and FDR trends across target FDR levels. These results demonstrate the robustness of AnnoGK in improving power while ensuring stricter FDR control relative to GhostKnockoff.

We also evaluated the performance of all methods using a binary annotation for each covariate: the annotation was set to 1 if the index was less than or equal to 60 and was set to 0 otherwise. As shown in Supplemental Figures S11, S12, and S13, both AnnoKn and AnnoGK exhibited robust performance regardless of the annotation type. Compared with Knockoffs and GhostKnockoff, both AnnoKn and AnnoGK consistently enhanced power and lower FDR under both binary and continuous annotations, and the extent of improvement was comparable across annotation types.

To validate AnnoKn and AnnoGK using real genotype data, we performed simulations using 10,000 UK Biobank (UKBB) participants. The LD matrix was estimated using either in-sample data or an external reference panel. As shown in Supplemental Figures S14 and S15 (and discussed in Supplementary Materials Section 5.2), AnnoKn and AnnoGK improved the power of detecting causal SNPs while simultaneously reducing FDR when using the in-sample LD matrix. Furthermore, when we used an external reference panel to estimate the LD matrix, AnnoGK exhibited greater robustness to LD mismatch between the reference panel and the summary statistics than GhostKnockoff, mitigating the FDR inflation caused by this mismatch.

Finally, we assessed the computational efficiency of AnnoKn and AnnoGK. We compared the running times of AnnoKn and AnnoKn-lite with those of AdaPT, Knockoffs, and AdaKn (Supplemental Figure S8). Under the one-dimensional annotation setting, AnnoKn exhibited computational efficiency comparable to AdaKn and became increasingly more efficient as the dimensionality of annotations increased. Moreover, AnnoKn-lite removed the need to search for the optimal $\lambda_0$ over a candidate set, substantially reducing computation time relative to AnnoKn. Consequently, AnnoKn-lite emerged as one of the most efficient annotation-informed variable selection methods. In addition, replacing Knockoffs and AnnoKn with their corresponding summary-statistics–based versions further reduced computation time (Supplemental Figure S16), highlighting the advantage of summary statistics over individual-level data in large-scale association studies.

\section{Real data applications}

\subsection{AnnoKn improved the identification of GTEx eGenes}

We applied AnnoKn to detect eGenes (genes whose expression level is affected by at least one independent expression quantitative trait locus (eQTL)) in the Genotype-Tissue Expression (GTEx) Project v8 \cite{gtex2020gtex}. Details of preprocessing are provided in Supplementary Materials Sections S7.5 and S7.6. The GTEx dataset contained tissue-specific gene expression measurements and whole-genome sequencing data, with the sample size for each tissue summarized in Supplemental Figure S17. In this study, we restricted our analysis to protein-coding genes from the ten GTEx tissues with the largest sample sizes and excluded genes with ten or fewer cis-SNPs. The numbers of genes with available expression data are summarized in Supplemental Table S1. For each gene, we included cis-SNPs located within the gene as well as those within 250 kb upstream and downstream.

Consistent with standard knockoff applications, we performed hierarchical clustering to group SNPs based on LD, using an absolute pairwise correlation threshold of $|r| > 0.5$. We then selected a single representative SNP from each cluster, specifically the SNP with the strongest marginal association with the phenotype (having the smallest $p$-value).

We obtained functional annotation information from stratified LD score regression (S-LDSC) \cite{finucane2015partitioning,gazal2024sldsc}, which provides 96-dimensional annotations for each SNP. We first compared the performance of AnnoKn with that of Knockoffs using three representative annotations whose associations with eQTLs have been well-studied: Promoter \cite{joehanes2017integrated, chandra2021promoter}, Enhancer \cite{gtex2020gtex, chen2016genetic}, and Transcription Factor Binding Site (TFBS) \cite{flynn2022transcription,natri2024cell}. Under $q$-value threshold of either 0.1 or 0.2, AnnoKn identified more eGenes than Knockoffs across all tissues evaluated, with average increase proportions of 12.8\% and 8.1\%, respectively (Table \ref{tab:annokn_results_combined_type1} and Supplemental Figure S18).

\begin{table}[htbp]
\centering
\renewcommand{\arraystretch}{1.2} 
\caption{Number of eGenes identified by Knockoffs and AnnoKn using three representative S-LDSC annotations across ten tissues under $q$-value thresholds of 0.1 or 0.2.}
\setlength{\tabcolsep}{3mm} 
\begin{tabular}{lcccc}
\hline
\textbf{Tissue} & \multicolumn{2}{c}{\textbf{Knockoffs}} & \multicolumn{2}{c}{\textbf{AnnoKn}} \\
\cline{2-3} \cline{4-5} 
  & \textbf{q = 0.1} & \textbf{q = 0.2} & \textbf{q = 0.1} & \textbf{q = 0.2} \\
\hline
Muscle Skeletal & 94 & 1,178 & 110 & 1,292 \\
Whole Blood & 75 & 1,229 & 81 & 1,322 \\
Skin Sun Exposed (Lower leg) & 96 & 1,409 & 102 & 1,519 \\
Artery Tibial & 115 & 1,424 & 138 & 1,524 \\
Adipose Subcutaneous & 92 & 1,336 & 106 & 1,473 \\
Thyroid & 119 & 1,568 & 143 & 1,661 \\
Nerve Tibial & 149 & 1,577 & 170 & 1,688 \\
Skin Not Sun Exposed (Suprapubic) & 99 & 1,420 & 110 & 1,556 \\
Lung & 104 & 1,266 & 112 & 1,387 \\
Esophagus Mucosa & 107 & 1,440 & 116 & 1,532 \\
\hline
\end{tabular}
\label{tab:annokn_results_combined_type1} 
\end{table}

These results indicate that AnnoKn is capable of identifying more eGenes by leveraging functional annotation information to prioritize eQTLs. We further evaluated whether the eGenes identified by AnnoKn were biologically more meaningful and more strongly enriched in tissue-relevant gene sets. As an external validation resource, we used the Jensen TISSUES database \cite{palasca2018tissues}, a tissue expression database integrating evidence from manually curated literature, proteomics, transcriptomics screens, and automated text mining. It assigned confidence scores to tissue-gene associations, enabling systematic comparison across diverse evidence sources. For each GTEx tissue, we selected the corresponding or most closely related tissue in Jensen TISSUES with substantial supporting evidence (Supplemental Table S1).

We assessed whether the eGenes identified by Knockoffs and AnnoKn were enriched in the tissue-specific Jensen gene sets relative to all candidate genes. Specifically, we conducted enrichment analysis using Enrichr \cite{kuleshov2016enrichr, xie2021gene}, with the background defined as all candidate genes for each tissue. This allowed us to compute a $p$-value representing the degree of overrepresentation of the identified eGenes in the match Jensen tissue, providing a quantitative measure of how well each method prioritized tissue-relevant genes. 

\begin{figure}[htbp]
  \centering
  \includegraphics[width=1\textwidth]{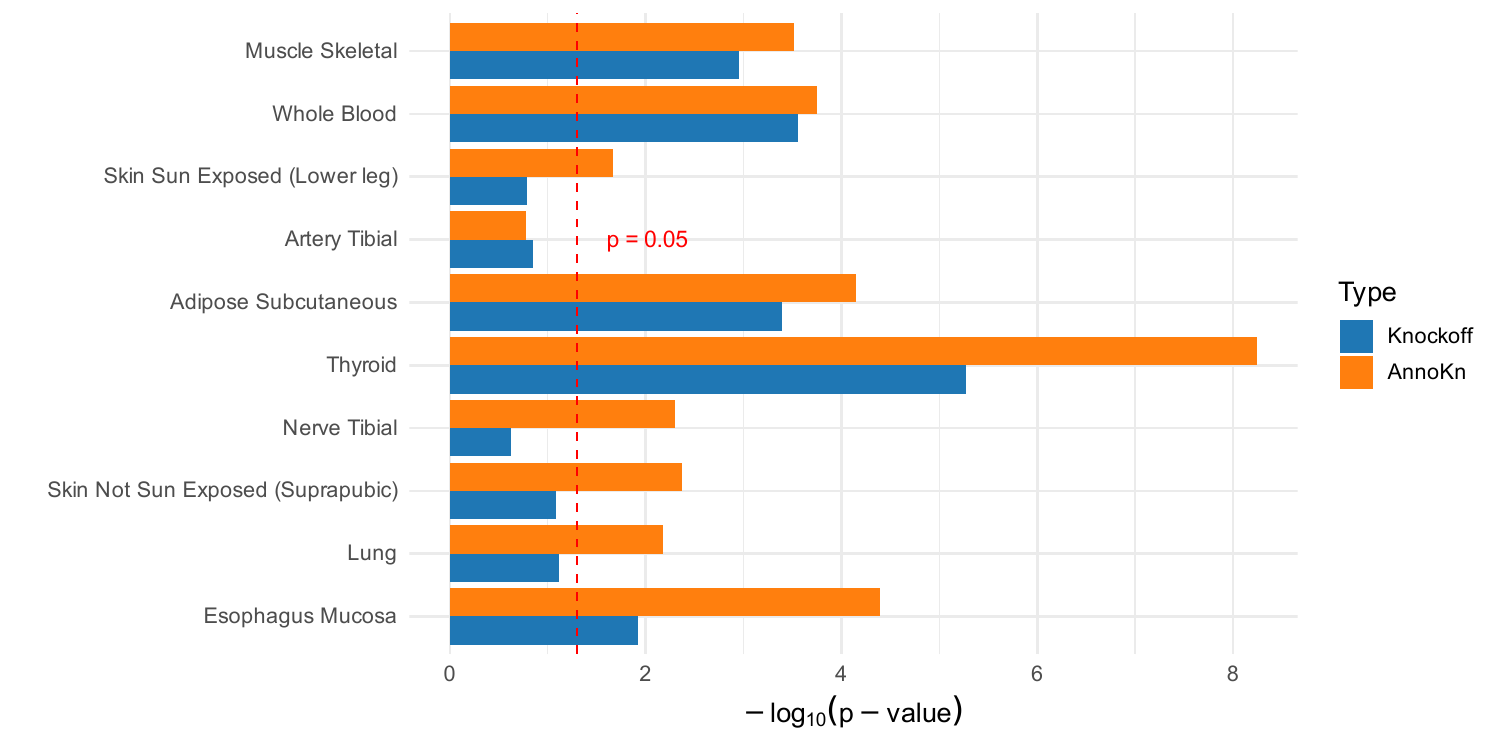}
  \caption{{\bf Comparison of enrichment of eGenes identified by Knockoffs and AnnoKn in GTEx tissues under the $q$-value threshold of 0.2.} We calculated the $-\log$($p$-value) for the overrepresentation of the identified tissue-specific eGenes in the matched Jensen tissue.}
  \label{GTEx_p-value}
\end{figure}

As shown in Figure \ref{GTEx_p-value}, eGenes identified by AnnoKn showed stronger enrichment in the Jensen database in 9 out of 10 GTEx tissues. The exception was Artery Tibial, where neither method produced significantly enriched eGenes. This was likely because the Jensen datasbase lacked a directly corresponding tissue; the closest available proxy, Coronary Artery, differed in biological properties and eGene profiles. At a significance threshold of $p < 0.05$, AnnoKn achieved significant enrichment in 9 out of 10 tissues except for Artery Tibial. In contrast, Knockoffs yielded significant enrichment in only 5 of the 10 tissues. These results demonstrate that AnnoKn captures increased and biologically more meaningful eGenes across diverse tissues than the standard knockoff framework, owing to its incorporation of functional annotations to prioritize informative eQTLs.

We also evaluated the estimated parameter $\lambda_l$ for each functional annotation. For Promoter, Enhancer, and TFBS, the annotations were binary variables, with a value of 1 indicating that a SNP was located within the corresponding functional region and 0 otherwise. To prioritize SNPs within functional regions as potential eQTLs, we aimed to assign them smaller adaptive lasso penalties, which corresponded to negative values of $\lambda_l$. To assess whether AnnoKn effectively prioritized functional genetic variants by learning reasonable parameter values, we examined the number of genes with strongly negative ($\lambda_l < -1$) or strongly positive ($\lambda_l > 1$) parameter estimates for each annotation. As shown in Supplemental Figure S19, the number of strong parameters was relatively small compared to the total number of protein-coding genes (hundreds out of approximately 18,000), indicating that the influence of functional annotations on eQTL identification was generally weak for most genes. Among these strong parameter values, the majority were negative. Across GTEx tissues, the average numbers of genes with strongly negative parameters ($\lambda_l < -1$) for Enhancer, Promoter, and TFBS were 480.0, 270.0, and 277.9, respectively. In contrast, the average numbers of genes with strongly positive parameters ($\lambda_l > 1$) for these annotations were much smaller: 13.1, 15.0, and 18.2, respectively. These results demonstrate that AnnoKn successfully leveraged functional annotations by assigning smaller penalty weights to biologically relevant covariates, which enhances the prioritization of eQTLs and identification of eGenes.

To further demonstrate the robustness of AnnoKn against the choice of annotations, we applied it to identify eGenes using all 96 functional annotations from the S-LDSC dataset. As presented in Supplemental Table S2 and Supplemental Figure S20 (and detailed in Supplementary Materials Section S6.1), incorporating the full annotation set enabled AnnoKn to identify more eGenes than Knockoffs at both $q$-value thresholds (13.7\% at 0.1 and 10.6\% at 0.2). These gains exceeded those obtained using only the three representative annotations, although the additional increase was relatively modest. This limited improvement could be explained by the fact that some additional annotations may be less relevant and contain highly redundant information. Furthermore, the eGenes identified by AnnoKn demonstrated stronger tissue-specific enrichment than those detected by Knockoffs (Supplemental Figure S21). These findings collectively suggest that the initially selected representative annotations already capture sufficient functional information, while incorporating all annotations provide additional, though comparatively modest, benefits.

\subsection{AnnoGK improves cross-ancestry GWASs}

To illustrate how GhostKnockoff could be informed by annotation information, we applied AnnoGK to leverage cross-ancestry information in two examples of GWASs aimed at identifying genetic loci for Height and Schizophrenia (SCZ), respectively. 

For the first study, we focused on a meta-analysis of Height involving European and African participants, combining data from multiple cohorts within the GIANT consortium \cite{yengo2022saturated} (details in Supplementary Materials Section S7.3). Separate Manhattan plots for both ancestries are available in Supplemental Figure S22. For both datasets, we retained only SNPs that overlapped with those in the 1KG Phase 3 reference panel \cite{consortium2015global}. We applied Popcorn \cite{brown2016transethnic} to estimate the transethnic genetic correlation. As summarized in Table \ref{heritability}, the genetic correlation of Height between European and African ancestry was 0.70 (s.e. $= 0.10$, $p$-value $= 0.002$), providing strong evidence that a large proportion of genetic variants associated with Height are shared between these two ancestries. 
\begin{table}[htbp]
\centering
\caption{Heritabilities and cross-ancestry genetic correlations of GWASs for Height and SCZ.}
\renewcommand{\arraystretch}{1.5}
\setlength{\tabcolsep}{3mm}
\begin{tabular}{cccc}
\hline
\textbf{Trait} & \textbf{Ancestry} & \textbf{Heritability (s.e.)} & \textbf{Genetic correlation (s.e.)} \\ \hline
\multirow{2}{*}{Height} & European & 0.38 (0.03) & \multirow{2}{*}{0.70 (0.10)}  \\ \cline{2-3}
& African & 0.33 (0.03) &   \\ \hline
\multirow{2}{*}{Schizophrenia} & European & 0.60 (0.06) & \multirow{2}{*}{0.75 (0.07)}  \\ \cline{2-3}
& East Asian & 0.62 (0.05) &   \\ \hline
\end{tabular}
\label{heritability}
\end{table}

We defined regions around significant genetic association signals ($p$-value $< 5\times 10^{-8}$) in the European Height GWAS, with all SNPs within 250 Kb upstream and downstream of each signal included in the same candidate region. Overlapping regions were merged until no further overlap remained. This procedure yielded 1,047 regions across all chromosomes (Supplemental Figure S23). As in the GTEx analysis, we performed hierarchical clustering to group SNPs with pairwise absolute correlation exceeding 0.5, and selected the SNP with the smallest marginal $p$-value as the representative SNP within each cluster. 

We then applied GhostKnockoff and AnnoGK to the European and African GWAS datasets to identify putative causal SNPs within these regions, using the 1KG reference panel. We defined significant regions as those containing at least one significant SNP. For each ancestry, we used the $-\log (p$-value$)$ of each SNP from the other ancestry as a continuous annotation, with larger annotation values indicating stronger evidence of association with Height in the other population. 

Under the FDR threshold of 0.1, GhostKnockoff identified 8,737 significant SNPs from 297 regions in the European GWAS and 3,376 significant SNPs from 132 regions in the African GWAS. After incorporating annotation information,  AnnoGK substantially increased both the number of significant SNPs and the number of significant regions for both ancestries (Table \ref{tab:height_gwas_signal}). We also compared the number of regions identified in both ancestries and observed a 44\% increase using AnnoGK. This gain exceeded the improvement in either single ancestry, suggesting that leveraging cross-ancestry genetic architecture not only enhances the number of discoveries but also improves the reproducibility of detected GWAS loci.
\begin{table}[htbp]
\centering
\caption{Number of causal SNPs (FDR threshold = 0.1) and regions with significant SNPs identified by GhostKnockoff and AnnoGK in Height GWASs.}
\label{tab:height_gwas_signal}
\renewcommand{\arraystretch}{1}
\setlength{\tabcolsep}{3mm}
\begin{tabular}{lccccc}
\toprule
\multirow{2}{*}{\textbf{Method}} & 
\multicolumn{2}{c}{\textbf{European GWAS}} & 
\multicolumn{2}{c}{\textbf{African GWAS}} & 
\multirow{2}{*}{\textbf{\shortstack{Common regions \\between two GWASs}}} \\ 
\cline{2-5} 
& \textbf{Region} & \textbf{SNP} & \textbf{Region} & \textbf{SNP} & \\ 
\midrule
GhostKnockoff & 297 & 8,737 & 132 & 3,376 & 88 \\
AnnoGK & 416 & 14,443 & 158 & 4,217 & 127 \\
\midrule
Increase & 40\% & 65\% & 20\% & 25\% & 44\% \\
\bottomrule
\end{tabular}
\end{table}

Similar to the GTEx analysis, we evaluated the estimated annotation parameter $\lambda_{l}$ for 416 significant regions identified by AnnoGK in the European GWAS. We expect SNPs with small $p$-values in the Arican ancestry to be less penalized in the adaptive Lasso model, thus $\lambda_{l}$ should be a negative value to prioritize SNPs with large annotations ($-\log (p$-values$)$). As shown in the Supplemental Figure S24, 78.8\% (328/416) of estimated $\lambda_l$ were negative, indicating that AnnoGK effectively learned the expected pattern of prioritizing SNPs with strong marginal associations in the other ancestry.

\begin{figure}[htbp]
    \centering
    \begin{subfigure}[b]{0.48\textwidth}  
        \centering
        \includegraphics[width=\textwidth]{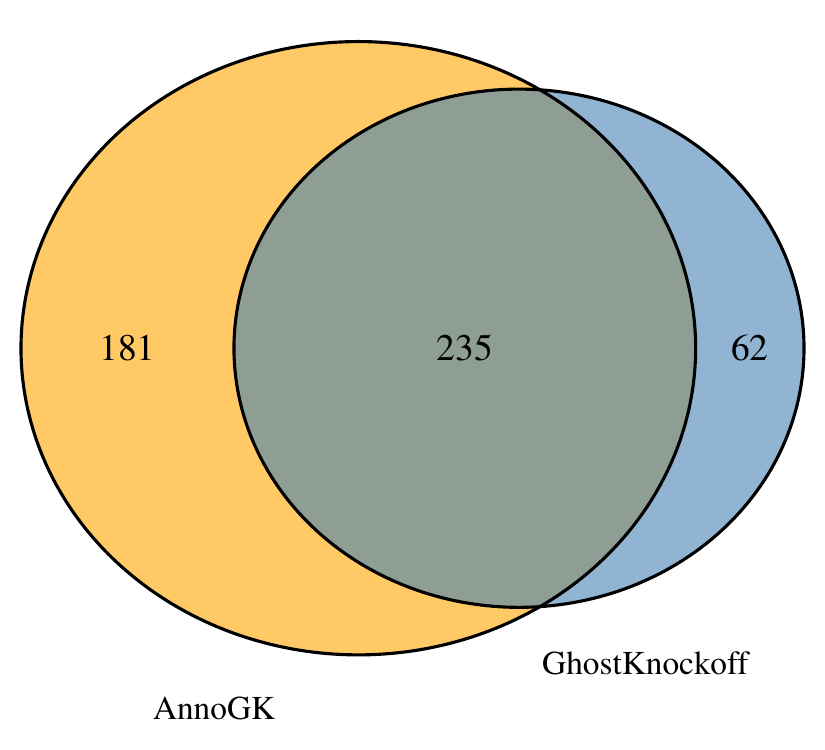}
        \label{Height_EUR}
    \end{subfigure}
    \hfill
    \begin{subfigure}[b]{0.48\textwidth}
        \centering
        \includegraphics[width=\textwidth]{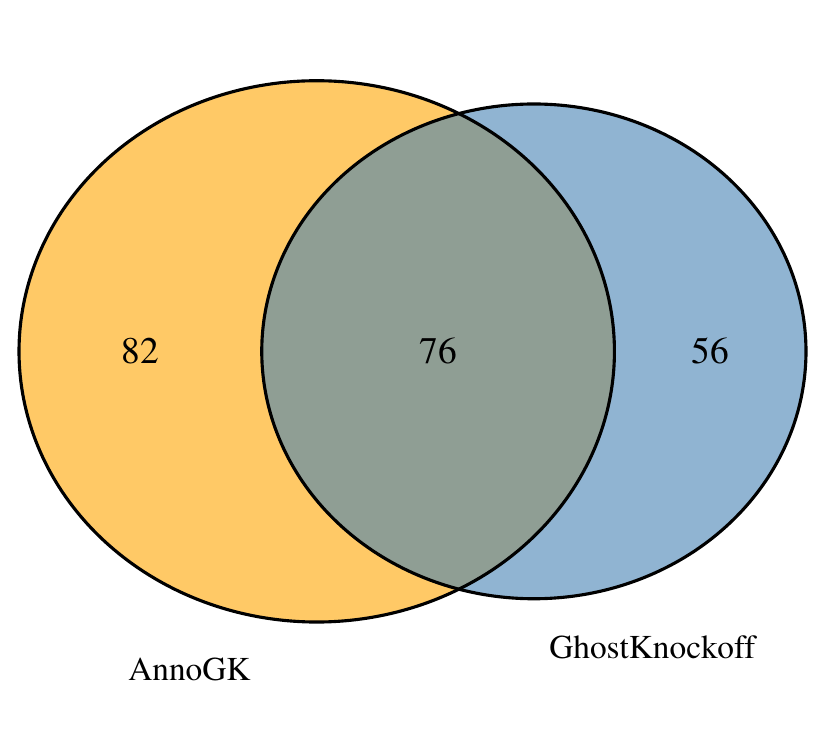}
        \label{Height_AFR}
    \end{subfigure}
    \caption{{\bf Venn plots for the significant regions detected by GhostKnockoff and AnnoGK in European (left) and African (right) Height GWASs.}}
    \label{Height_GWAS}
\end{figure}

Finally, we compared the overlap of significant regions identified by GhostKnockoff and AnnoGK in the European and African Height GWASs. As shown in Figure \ref{Height_GWAS}, most of the regions detected by GhostKnockoff were also recovered by AnnoGK. However, not all regions identified by GhostKnockoff were identified by AnnoGK. This discrepancy could be partly explained by two factors. First, the inherent randomness in the knockoff construction introduced variability in discoveries across runs. Second, AnnoGK may fail to detect a region when functional annotations provide insufficient support for it, causing the corresponding signal to be deprioritized in the annotation-informed procedure.

For the second example, we applied AnnoGK to a meta-analysis of SCZ involving individuals from four ancestries: European (EUR), East Asian (EAS), African American (AFR), and Latino (LAT) \cite{trubetskoy2022mapping} (see Supplemental Figure S25 and Supplementary Materials Section S7.4 for details). In this analysis, we focused primarily on two ancestries with relatively large sample sizes: EUR and EAS, while using summary statistics from the remaining ancestries as annotation information. As summarized in Table \ref{heritability}, the genetic correlation of SCZ between EUR and EAS was very significant ($p$-value $= 0.0003$), indicating that borrowing information across ancestry to detect causal SNPs for SCZ is meaningful.

Following the same procedure as in the Height GWAS application, we obtained 169 regions based on summary statistics from the European SCZ GWAS (Supplemental Figure S26). One region was located in the Major Histocompatibility Complex (MHC) region, and we removed it because of the complicated LD structure, a common practice in GWASs and other downstream analyses \cite{uffelmann2021genome}. We then compared the performance of AnnoGK with GhostKnockoff in detecting regions containing significant causal SNPs for the EUR and EAS SCZ GWASs, using the 1KG dataset as the reference panel. For each ancestry, we used the $-\log (p$-values$)$ from all three other ancestries as multiple-dimensional continuous annotations. 

Taking into account the inherent randomness of the knockoff procedure \cite{ren2024derandomised}, the sets of selected covariates can vary across repeated runs. This variability is expected to be larger for the SCZ GWASs, where the associations are generally weaker than those for Height. To mitigate this issue, we repeated this analysis for 10 independent knockoff copies. When aggregating results across all repetitions, incorporating information from other ancestries through AnnoGK increased the total number of regions containing significant SNPs under an FDR threshold of 0.1 by 33\% and 23\% in the EUR and EAS studies, respectively (Table \ref{tab:SCZ_gwas_signal}). In addition, annotation-informed variable selection increased the number of regions identified in both ancestries from 2 to 12. In other words, after prioritizing candidate SNPs using marginal associations from other ancestries, 75\% (12/16) of significant regions in the EAS GWAS were reproduced in the EUR GWAS.
\begin{table}[htbp]
\centering
\caption{Number of causal SNPs (FDR threshold = 0.1) and regions with significant SNPs identified by GhostKnockoff and AnnoGK in SCZ GWASs for EUR and EAS ancestries, aggregated across 10 knockoff repetitions.}
\label{tab:SCZ_gwas_signal}
\renewcommand{\arraystretch}{1}
\setlength{\tabcolsep}{4mm}
\begin{tabular}{lccccc}
\toprule
\multirow{2}{*}{\textbf{Method}} & 
\multicolumn{2}{c}{\textbf{EUR GWAS}} & 
\multicolumn{2}{c}{\textbf{EAS GWAS}} & 
\multirow{2}{*}{\textbf{\shortstack{Common regions \\between two GWASs}}} \\ 
\cline{2-5} 
& \textbf{Region} & \textbf{SNP} & \textbf{Region} & \textbf{SNP} & \\ 
\midrule
GhostKnockoff & 40 & 907 & 13 & 198 & 2 \\
AnnoGK & 53 & 1,243 & 16 & 317 & 12 \\
\midrule
Increase & 33\% & 37\% & 23\% & 60\% & 500\% \\
\bottomrule
\end{tabular}
\end{table}

To evaluate the reproducibility across knockoff repetitions, we examined how many regions were identified in at least a given percentage of the repetitions. As shown in Supplemental Figure S27, AnnoGK achieved higher numbers of regions than GhostKnockoff across all detection frequency thresholds. For example, at a detection frequency threshold of at least 0.5, AnnoGK identified 6 and 1 reproducible regions in the EUR and EAS GWASs, respectively, whereas GhostKnockoff identified only 2 and 0. Supplemental Figure S28 shows similar improvements in the number of reproducible SNPs. These results demonstrate that AnnoGK improves the stability of knockoff-based variable selection by leveraging annotation information to better prioritize causal SNPs.

\section{Discussion}
This paper introduced AnnoKn, a novel framework for incorporating functional annotations into the knockoff procedure. This approach distinguishes itself from existing methods by integrating the adaptive Lasso with knockoffs through a Bayesian framework, enabling controlled variable selection while simultaneously leveraging functional annotations to prioritize covariates within a unified process. We further extended AnnoKn to AnnoGK, which operates on summary statistics and a reference panel for estimating genetic correlations. This extension substantially enhances the practical utility of annotation-informed knockoffs in large-scale genomic applications. 

Through simulations and real data analyses, we demonstrated the benefits of AnnoKn in improving the power of variable selection by incorporating prior information on these covariates. We highlighted two practical and impactful applications of AnnoKn. First, AnnoKn can prioritize causal genetic variants using functional annotations or multi-omics data, as illustrated in our GTEx eGene analysis. Second, AnnoKn provides a principled way to borrowing information across ancestries, thereby improving association studies in underrepresented populations where limited sample sizes often reduce statistical power. 

A key strength of AnnoKn and AnnoGK lies in their ability to efficiently handle high-dimensional annotation data and to adaptively identify the annotations that contribute most to the model. This capability removes the need for subjectively specifying annotation weights or conducting a separate preprocessing step for annotation selection prior to the main analysis. Another advantage is that the objective function has favorable properties, ensuring stability and computational efficiency of AnnoKn. Across all simulation settings and real data applications, we did not observe any failures of the AnnoKn and AnnoGK algorithms, whereas AdaKn occasionally encountered convergence issues and failed to produce valid results. In addition, although the knockoff procedure is inherently random due to the generation of knockoff copies. AnnoKn is expected to improve stability of the knockoff method by leveraging annotations as consistent information to help prioritize candidate covariates.

The flexible modeling framework of AnnoKn also opens several directions for future extensions and methodological developments. First, the current model links the penalty terms for adaptive Lasso to a linear combination of annotations, limiting the method to capture linear relationships in the annotation data. While this assumption is generally sufficient, given that most commonly used annotations have clear directional interpretations (e.g., an indicator of functional regions or smaller $p$-values representing stronger associations), it would still be worth exploring more flexible link functions to accommodate increasingly rich and complex annotation resources in modern biomedicine and other research fields. 

Secondly, the use of AnnoKn to leverage cross-ancestry GWAS information may appear similar to other meta-analytic variable selection approaches, particularly cross-ancestry fine-mapping methods \cite{yuan2024fine, lapierre2021identifying, cai2023xmap}. However, we found it difficult to directly compare results from AnnoKn with these methods for several reasons. (1) Most cross-ancestry fine-mapping methods are Bayesian and define discoveries in a different way, focusing on identifying sets of highly correlated SNPs whose cumulative posterior probability exceeds a predefined threshold. In contrast, AnnoKn first clusters highly correlated covariates into groups and then performs FDR-controlled variable selection on the representatives. The outputs are therefore not directly comparable to posterior inclusion probabilities (PIPs). (2) Although AnnoKn borrows information from multiple ancestries, its primary goal is to improve inference in the primary study using external information. This differs from cross-ancestry fine-mapping, which aims to uncover both shared and ancestry-specific genetic variants. Developing systematic approaches to compare AnnoKn with cross-ancestry fine-mapping methods remains an interesting direction for future research.

In summary, AnnoKn provides a reliable and principled framework to incorporate prior knowledge about covariates into variable selection for high-dimensional nonlinear models. This methodology addresses a fundamental statistical challenge and has broad applicability across a wide range of scientific domains  beyond statistical genetics.

\section{Acknowledgments}
Supported in part by NIH grants U24 HG012108 and U01 HG013840.

\section{Declaration of interests}

The authors declare no competing interests.

\section{Web resources}
We used the UKBB genotype data obtained from the resource \url{https://www.ukbiobank.ac.uk/}. The GTEx dataset was obtained at \url{https://www.ncbi.nlm.nih.gov/projects/gap/cgi-bin/study.cgi?study_id=phs000424.v8.p2}. The 1KG reference panel was available at \url{https://www.cog-genomics.org/plink/2.0/resources}. Summary statistics for the Height GWAS could be found at GIANT consortium (\url{https://giant-consortium.web.broadinstitute.org/index.php/GIANT_consortium_data_files}). Summary statistics for SCZ were provided by the Psychiatric Genomics Consortium (PGC) at \url{https://figshare.com/articles/dataset/scz2022/19426775}. The Jensen tissue expression database was available at \url{http://tissues.jensenlab.org/}.

\bibliographystyle{unsrt}
\bibliography{ref.bib}

\end{document}